\documentclass[10pt]{iopart}
\usepackage{iopams,setstack,bm,cite,graphicx,mathptmx,url,color}

\newcommand{\Gal}[1]{\mathbb{F}_{#1}}
\newcommand{\XOR}{\mbox{\textsc{xor}}}

\newcommand{\prim}{\sigma}
\newcommand{\openone}{\leavevmode\hbox{\small1\normalsize\kern-.33em1}}

\eqnobysec

\begin{document}

\title{Symmetric discrete coherent states for $n$ qubits}

\author{C Mu\~{n}oz$^{1}$, A B Klimov$^1$ and L L S\'{a}nchez-Soto$^{2}$}

\address{$^1$ Departamento de F\'{\i}sica, Universidad de Guadalajara,
44420~Guadalajara, Jalisco, Mexico}

\address{$^2$ Departamento de \'Optica, 
Facultad de F\'{\i}sica, Universidad Complutense, 
28040~Madrid, Spain}

\date{\today}

\begin{abstract}
  We put forward a method of constructing discrete coherent states for $n$
  qubits. After establishing appropriate displacement operators, the
  coherent states appear as displaced versions of a fiducial vector
  that is fixed by imposing a number of natural symmetry requirements on its $Q$
  function.  Using these coherent states we establish a partial order
  in the discrete phase space, which allows us to  picture some
  $n$-qubit states as apparent distributions. We also analyze
  correlations in terms of sums of squared $Q$ functions.
\end{abstract}


\section{Introduction}

Coherent states (CS), first introduced by
Schr\"{o}dinger~\cite{Schrodinger:1926uq}, are of paramount
significance for modern physical theories, as they are quantum states
that follow classical trajectories. In quantum optics, CS were
popularized by Glauber~\cite{Glauber:1963kx,Glauber:1963ys} for the
description of correlation properties of a single-mode radiation
field, where the Weyl-Heisenberg group emerges as a hallmark of
noncommutativity~\cite{Binz:2008oq}.

Although generalizations along several directions have been considered
(see references ~\cite{Klauder:1985lr,Zhang:1990fk,Gazeau:2009kx} for
reviews), nowadays it seems  indisputable that the pioneer work
of Perelomov~\cite{Perelomov:1986zr} paved the way to extend the
notion of CS to quantum systems with dynamical symmetry group $G$. In
this approach, CS appear as orbits of a certain fiducial state under a
unitary irreducible representation of $G$ acting in the corresponding
Hilbert space.

This fiducial state is chosen to have maximal isotropy subgroup $H$,
which ultimately leads to ``maximal classicality''. Under these
circumstances, the displacement operators, transforming CS among
themselves, are labeled by points in the manifold $M=G/H$. For many
physical models, $M$ can be equipped with an irreducible symplectic
structure~\cite{Kostant:1970uq,Kirillov:1976ly,Souriau:1970lr,Guillemin:1990yq},
so it can be considered as the phase space of a classical dynamical
system. In other words, there is a one-to-one correspondence between
CS and points of the classical phase space.

The crucial point of this construction is that the Hilbert space is
irreducible under the action of $G$. This is especially clear for the
symmetric representations of the unitary groups $SU(n)$, when the
classical phase space is a $(2n-2)$-dimensional
sphere~\cite{Wybourne:1974ul}. For the discrete counterparts, even in
the physically (but not mathematically!) simple $n$-qubit case, the
symmetric subspace is only a very small portion of the whole
$2^{n}$-dimensional Hilbert space. Nonetheless, one can define a
natural set of CS, constructed in a similar way as in the continuous
case: acting on a fiducial state with discrete displacements; i.e.,
unitary operators labeled by elements of two discrete
sets~\cite{Schwinger:1960a}.  These two sets can be organized in a
discrete $2^{n}\times 2^{n}$ grid, on which a specific discrete
geometry (including symplectic operations) can be introduced, so that
such a grid turns out to be a \emph{bona fide} discrete phase
space~\cite{Wootters:1987qf,Galetti:1988lq,Galetti:1992ve,
Kasperkovitz:1994,Galetti:1996rr,Gibbons:2004bh,Vourdas:2004,
Wootters:2004,Vourdas:2007qe}.

Although the points of the discrete phase space label again CS, there
is still an essential difference with the continuous case: the choice
of the fiducial state. For continuous symmetry groups, the standard
choice corresponds to an extreme state of the representation, such as
the vacuum or the lowest/highest weight state. In the discrete case,
the nature of the unitary displacements prevents such a simple notion
and different possibilities have been
discussed so far~\cite{Munoz:2009wd,Klimov:2009eu}.

In this paper, we take the fiducial state as a spin CS and impose that
its associated $Q$ function fulfills reasonable symmetry
conditions. This not only solves the problem, but allows us to use the
system of CS to impose a partial order in phase space, which helps to
recognize states pictured as distributions. Finally, we briefly
speculate about detecting quantum correlations through the sum of
squared $Q$ functions.

\section{Discrete phase space}

A qubit is a two-dimensional quantum system, with Hilbert space
isomorphic to $\mathbb{C}^{2}$. It is customary to choose two
normalized orthogonal states, $\{ | 0 \rangle, | 1 \rangle \}$, as a
computational basis. The unitary operators
\begin{equation}
  \sigma_{z} = | 0 \rangle \langle 0 | - | 1 \rangle \langle 1 | \, , 
  \qquad \qquad 
  \sigma_{x} = | 0 \rangle \langle 1 | + | 1 \rangle \langle 0 | \, .
  \label{sigmas}
\end{equation}
generate the Pauli group $\mathcal{P}_{1}$ under matrix
multiplication~\cite{Chuang:2000fk}.

For $n$ qubits, the Hilbert space is the tensor product
$\mathbb{C}^{2} \otimes \cdots \otimes \mathbb{C}^{2} =
\mathbb{C}^{2^{n}}$. A compact way of labeling both states and
elements of the corresponding Pauli group $ \mathcal{P}_{n}$ consists
in using the finite field $\Gal{2^{n}}$~\cite{Lidl:1986fk}. This can
be considered as a linear space spanned by an abstract basis $\{
\theta_{1}, \ldots , \theta_{n} \}$, so that given a field element
$\alpha $ (henceforth, field elements will be denoted by Greek
letters) the expansion
\begin{equation}
  \alpha = \sum_{i=1}^{n} a_{i} \,\theta_{i} \, , 
  \qquad 
  a_{i}\in \mathbb{Z}_{2} \, ,  
  \label{alpha}
\end{equation}
allows us the identification $\alpha \Leftrightarrow (a_{1}, \ldots,
a_{n})$. Moreover, the basis can be chosen to be orthonormal with
respect to the trace operation (the self-dual basis); that is,
\begin{equation}
  \tr ( \theta_{i} \, \theta_{j}) = \delta_{ij} \, ,
\end{equation}
where $\tr (\alpha )= \alpha +\alpha^{2} + \ldots + \alpha
^{2^{n-1}}$, which actually maps $\mathbb{F}_{2^{n}} \mapsto
\mathbb{Z}_{2}$. In this way, we associate each qubit with a
particular element of the self-dual basis: qubit$_{i} \Leftrightarrow
\theta_{i}$.

The generalized Pauli group $\mathcal{P}_{n}$ is generated now by the
operators 
  \begin{equation}
    Z_{\alpha}  =  \sum_{\lambda} \chi ( \alpha \lambda  ) \,
    | \lambda \rangle \langle \lambda | \, ,
    \qquad \qquad 
    X_{\beta} = \sum_{\lambda}
    |  \lambda + \beta \rangle \langle \lambda | \, .
    \label{XZgf} \\
  \end{equation}
  Here the additive characters $\chi $ are defined as $\chi (\alpha )
  = \exp [ i \pi \tr ( \alpha ) ]$ and $| \lambda \rangle$  is an
  orthonormal basis in the Hilbert space of the system.
  Operationally, the elements of the basis can be labeled by powers of
  a primitive element (i.e., a root of the primitive polynomial), and
  read $\{|0 \rangle, \, |\prim \rangle, \ldots, \, |\prim^{2^n-1} =
  1 \rangle \} $.  One can verify that
  \begin{equation}
    Z_{\alpha} X_{\beta} = \chi ( \alpha \beta ) \, X_{\beta} Z_{\alpha}\, ,
  \end{equation}
  which is the discrete counterpart of the Weyl-Heisenberg algebra for
  continuous variables~\cite{Binz:2008oq}. 

The operators (\ref{XZgf}) can be factorized into tensor products of
powers of single-particle Pauli operators. This factorization can be
carried out by mapping each element of $\Gal{2^n}$ onto an ordered
set of natural numbers according to
\begin{equation}
    Z_{\alpha}  = \sigma_{z}^{a_{1}} \otimes \ldots \otimes
    \sigma_{z}^{a_{n}},
    \qquad \qquad
    X_{\beta} = \sigma_{x}^{b_{1}} \otimes \ldots \otimes
    \sigma_{x}^{b_{n}} \, ,
\end{equation}
 where $a_{i} = \tr ( \alpha \theta_{i} )$ and $b_{i} = \tr ( \beta
 \theta_{i} ) $ are the corresponding expansion coefficients for
 $\alpha$ and $\beta$ in the self-dual basis. Moreover, they are
 related through the finite Fourier transform~\cite{Klimov:2005}
 \begin{equation}
    \mathcal{F}=\frac{1}{\sqrt{2^{n}}}\sum_{\lambda ,\lambda^{\prime}}
    \chi (\lambda \,\lambda^{\prime})\, 
    |\lambda \rangle \langle \lambda^{\prime}| \, ,
    \label{FTcomp}
 \end{equation}
 so that $X_{\mu}=\mathcal{F}Z_{\mu}\,\mathcal{F}$.

 We next recall~\cite{Gibbons:2004bh} that the grid defining the phase
 space for $n$ qubits can be appropriately labeled by the discrete
 points $(\alpha, \beta)$, which are precisely the indices of the
 operators $Z_{\alpha}$ and $ X_{\beta}$: $\alpha$ is the
 ``horizontal'' axis and $\beta$ the ``vertical'' one. In this grid
 one can introduce the concept of straight lines (also called rays),
 which possess the same properties as in the continuous case.  It is
 worth noting that the monomials labeled by points of the same ray,
 $\{ (\alpha ,\mu \alpha ) \}$ commute with each other, so that one
 can establish a correspondence between eigenstates of such commuting
 sets and states (actually bases) in the Hilbert
 space~\cite{Gibbons:2004bh}.

Following our program, we introduce the set of displacements 
\begin{equation}
  D ( \alpha ,\beta )  =  e^{i\Phi (\alpha ,\beta )}  \, 
  Z_{\alpha}  X_{\beta} \, ,  
  \label{Dop} 
\end{equation}
where $\Phi ( \alpha, \beta)$ is a phase required to avoid plugging
extra factors when acting with $D$. One can immediately check that
\begin{equation}
  D ( \alpha ,\beta)  D^{\dag} ( \alpha ,\beta ) = \openone, 
  \quad 
  D^{\dag} ( \alpha ,\beta ) = D ( \alpha ,\beta ) \, ,
\end{equation}
so they are unitary and Hermitian. They also constitute a
complete trace-orthonormal set
\begin{equation}
  \Tr [ D( \alpha,\beta) \, D( \alpha^{\prime} , \beta^{\prime} ) ]
  =2^{n}\delta_{\alpha \alpha^{\prime}}\delta_{\beta \beta^{\prime}}\,.
\end{equation}
These operators act multiplicatively on the monomials (\ref{XZgf}),
thus shifting phase-space points according to
\begin{equation}
  (\alpha ,\beta )\overset{D(\alpha^{\prime},\beta^{\prime})}{\mapsto}
  (\alpha +\alpha^{\prime},\beta +\beta^{\prime})\,,
\end{equation}
which justifies their designation.

Finally, for later purposes, we touch on a pair of symplectic
operations ($z$- and $x$-rotations) that transform rays into rays
according to
\begin{equation}
  P_{\mu}Z_{\alpha}P_{\mu}^{\dagger} \propto 
  Z_{\alpha}X_{\mu \alpha}\, , 
  \qquad \qquad 
  Q_{\nu} X_{\beta}Q_{\nu}^{\dagger} \propto 
  Z_{\nu \beta}X_{\beta} \, .
\end{equation}
The symbol $\propto $ indicates equality except for a phase. Both
$P_{\mu}$ and $Q_{\nu}$ are unitary operators, with $[ P_{\mu}, 
X_{\nu} ] = [ Q_{\nu}, Z_{\mu}]=0$, and can be written as
\begin{equation}
  P_{\mu} = \sum_{\lambda} c_{\lambda ,\mu} 
  |\widetilde{\lambda}\rangle \langle \widetilde{\lambda}|\, ,
  \qquad  \qquad
  Q_{\nu} = \sum_{\lambda} c_{\lambda,\nu} 
  |\lambda \rangle \langle \lambda |  \, ,
  \label{V}
\end{equation}
where $|\lambda \rangle $ are the eigenstates of $Z_{\alpha}$ and $|
\widetilde{\lambda}\rangle $ of $X_{\beta}$. The coefficients
$c_{\lambda,\nu}$ fulfill the recurrence relation
\begin{equation}
  c_{\lambda +\alpha ,\nu} = c_{\alpha ,\nu} c_{\lambda ,\nu}\,
  \chi (\nu \alpha \lambda ),
  \qquad \qquad
  c_{0,\nu}=1 \, , 
  \label{cc}
\end{equation}
whose explicit solution can be found in reference~\cite{Klimov:2006vn}
and it is unimportant for the rest of the paper.

We associate an eigenstate $|\psi_{0}\rangle $ of $Z_{\alpha}$ with
the horizontal axis and immediately obtain that the state associated
with the ray $\beta =\mu \alpha $ is $P_{\mu} |\psi_{0}\rangle$, while
the vertical axis is associated with $
\mathcal{F}|\psi_{0}\rangle$~\cite{Klimov:2007} . Any other straight
line, parallel to a given ray, but crossing the axis $\beta $ at the
point $\beta =\xi$, corresponds to the state $X_{\xi} P_{\mu} |
\psi_{0} \rangle$.

Using phase-space coordinates these $z$- and $x$-rotations can be
interpreted as
\begin{equation}
  (\alpha ,\beta ) \overset{P_{\mu}}{\mapsto}
  (\alpha ,\beta +\mu \alpha ) \, , 
  \qquad \qquad
  (\alpha ,\beta ) \overset{Q_{\nu}}{\mapsto}
  (\alpha +\nu \beta ,\beta ) \, .  
  \label{Pab}
\end{equation}
It is clear that these two transformations are conjugate each other.

\section{Discrete coherent states for $n$ qubits}

According to the conventional approach, we define discrete CS
$|\alpha ,\beta \rangle $, labeled by phase-space points
$(\alpha,\beta )$,  as the displacements of the fiducial state 
$|\Psi_{0}\rangle$~\cite{Galetti:1996rr}:
\begin{equation}
  |\alpha ,\beta \rangle =D(\alpha ,\beta )|\Psi_{0}\rangle \, .
  \label{discretecs}
\end{equation}
The state $|\Psi_{0}\rangle $ can be chosen in several
ways~\cite{Munoz:2009wd}. Here, for reasons that will be apparent
soon, we take $|\Psi_{0}\rangle $ as a product of identical qubit
states:
\begin{equation}
  |\Psi_{0}\rangle =|\vartheta ,\varphi \rangle_{1}\otimes \cdots \otimes
  |\vartheta ,\varphi \rangle_{n} \, ,  
  \label{cs tp}
\end{equation}
where
\begin{equation}
  |\vartheta ,\varphi \rangle_{j} = e^{i\varphi /2} 
  \sin \left( \frac{\vartheta }{2}\right) |1\rangle_{j} 
  + e^{-i\varphi /2} 
  \cos \left ( \frac{\vartheta}{2} \right) |0\rangle_{j} \, ,
\end{equation}
and the angles $(\vartheta ,\varphi )$ parametrize the Bloch
sphere. The state (\ref{cs tp}) is invariant under permutation of the
qubit indices and thus can be expanded as~\cite{Munoz:2009wd}
\begin{equation}
  |\Psi_{0}\rangle \equiv |\xi \rangle =\frac{1}{(1+|\xi |^{2})^{n/2}}
  \sum_{k=0}^{n}\sqrt{\frac{n!}{k!(n-k)!}}\,\xi^{k}|k,n\rangle \, ,
  \label{su2coherentstate}
\end{equation}
with $\xi =e^{i\varphi}\,\tan \vartheta /2$.  The basis $\{|k,n\rangle
\,:\,k=0,\ldots ,n\}$ are the Dicke states
\begin{equation}
  |k,n\rangle =\sqrt{\frac{k!(n-k)!}{n!}}\sum_{k=0}^{n}
  \mathcal{P}_{k}| (|1\rangle_{1}\ldots |1 \rangle_{k}
  | 0 \rangle_{k+1}\ldots | 0\rangle_{n}) \, .  
  \label{DS}
\end{equation}
Here, $\{\mathcal{P}_{k}\}$ denotes the complete set of all the
possible qubit permutations.

In field notation, the state (\ref{su2coherentstate}) can be compactly
expressed as
\begin{equation}
  | \xi \rangle =\frac{1}{(1+|\xi |^{2})^{n/2}}
  \sum_{\kappa}\xi^{h(\kappa)}|\kappa \rangle \, ,  
  \label{ksi_h}
\end{equation}
where the function $h(\kappa )$ counts the number of nonzero
coefficients $k_{j}$ in the expansion of $\kappa $ in the field basis
(see the appendix for a brief account of its properties).

Finally, notice that one might think in imposing that the states $|
\xi\rangle$ are eigenstates of the Fourier
operator~\cite{Mehta:1987dp,Ruzzi:2006ai} (much as the vacum is for
continuous variables).  Since $\mathcal{F}^{2} = \openone$, this is
tantamount to
\begin{equation}
  \label{eq:Foureigen} 
  \mathcal{F} | \xi \rangle = \pm  | \xi \rangle \, ,
\end{equation}
which leads to two possible candidates $\xi_{\pm} = \pm \sqrt{2}-1$
and all the qubits pointing in the same direction. However, we prefer
to follow an alternative route to fix the possible values of $\xi$.

\subsection{$P$-function}

Let us look for an expansion of the density matrix of the form
\begin{equation}
  \varrho =\sum_{\alpha ,\beta} P(\alpha ,\beta )
  |\alpha ,\beta \rangle \langle \alpha ,\beta | \, ,
\end{equation}
which is the analogous to the Glauber-Sudarshan $P$-function for
continuous variables~\cite{Ruzzi:2005sd,Marchiolli:2005rm}. It is not
difficult to check that the function $P (\alpha ,\beta )$ may be
recast as
\begin{equation}
  P(\alpha ,\beta )= \Tr [ \varrho \Delta ( \alpha ,\beta ) ]\, ,
\end{equation} 
where $\Tr$ (with capital T to distinguish it from $\tr$, which is
the trace in the field) stands for the ordinary trace
operation in Hilbert space.  The kernel $\Delta (\alpha ,\beta )$
reads
\begin{equation}
  \Delta (\alpha , \beta )=\frac{1}{2^{2n}} \sum_{\gamma ,\delta}
  \chi (\alpha \delta +\beta \gamma ) \, 
  \langle \xi | D(\gamma ,\delta |\xi \rangle^{-1} 
  D(\gamma ,\delta ) \, .
\end{equation}
This clearly shows that the $P$-function is nonsingular only when $\langle
\xi | D(\gamma ,\delta ) |\xi \rangle^{-1}$ exists.

Since the state $|\xi \rangle $ is factorized into single-qubit
states, we have
\begin{equation}
  \langle \xi |D(\gamma ,\delta ) |\xi \rangle \propto 
  \prod_{i=1}^{n}\langle \xi^{(1)}| \sigma_{z}^{g_{i}} 
  \sigma_{x}^{d_{i}}|\xi^{(1)}\rangle \, ,
\end{equation}
where $g_{i}, d_{i}$ are the expansion coefficients of $\gamma $ and $\delta $
in the self-dual basis and 
\begin{equation}
|\xi^{(1)}\rangle =\frac{1}{\sqrt{1+|\xi |^{2}}} 
 (|0\rangle +\xi |1\rangle )\, .  
\label{eq:xi1}
\end{equation}
Using equation (\ref{ksi_h}) one can find 
\begin{eqnarray}
  \fl 
  \langle \xi | D (\gamma ,\delta )|\xi \rangle &\propto &
  \left( \frac{1-|\xi |^{2}}{1+|\xi |^{2}} 
  \right)^{[h(\gamma )-h(\delta )+h(\gamma +\delta )]/2}  \nonumber \\
  & \times & \left ( \frac{\xi +\xi^{\ast}}{1+|\xi |^{2}}
  \right )^{[h(\delta)-h(\gamma )+h(\gamma +\delta )]/2}
  \left ( \frac{\xi -\xi^{\ast}}{1+|\xi|^{2}}
  \right)^{[h(\gamma )+h(\delta )-h(\gamma +\delta )]/2} \, .
  \label{D_ksi}
\end{eqnarray}
This obviously rules out some values of $\xi $ for which $P$ is singular.

As an illustrative example, consider the case of a single qubit. Then, one
finds that 
\begin{equation}
  \fl 
  P( a,b) =\frac{1}{4}+ \frac{1+ | \xi |^{2}}{4} 
  \left[ ( -1)^{b} 
    \frac{\Tr ( \varrho \sigma_{z})}{1-| \xi |^{2}} +
    ( -1)^{a}
    \frac{\Tr ( \varrho\sigma_{x})}{\xi +\xi^{\ast}}
    + ( -1)^{a+b} 
    \frac{\Tr ( \varrho \sigma_{z} \sigma_{x})}{\xi-\xi^{\ast}} 
  \right ] \, ,  
  \label{P1a}
\end{equation}
where now $a, b \in \mathbb{Z}_{2}$. This function is singular when $\xi $ is real,
imaginary, and when $| \xi |^{2}=1$, i.e. on the equator of the Bloch
sphere. In particular, the eigenstates of the discrete Fourier transform,
when $\xi_{\pm} = \sqrt{2}\pm 1$, do not lead to the faithful expansion on
CS for qubits, contrary to what one could expect.

\subsection{$Q$-function}

In our search for determining the values of $\xi $, we next look at
the $Q$-function, defined in complete analogy with its continuous
counterpart, namely
\begin{equation}
  Q_{\varrho} (\alpha , \beta ) = 
  \langle \alpha ,\beta |\varrho |\alpha ,\beta \rangle \, ,  
  \label{eq:defQ}
\end{equation}
which satisfies 
\begin{equation}
\sum_{\alpha ,\beta} Q_{\varrho} (\alpha ,\beta )=2^{n} \, .  
\label{eq:defSQ}
\end{equation}
Let us impose the maximal symmetry conditions admissible on the
$Q$-function for the fiducial state $|\xi \rangle$:
\begin{description}
\item[C1] .- $Q_{|\xi \rangle} ( \alpha ,\beta )$ is symmetric under
  axis permutations.

\item[C2] .- The values of $Q_{|\xi \rangle}(\alpha ,\beta )$ can be
  obtained from $Q_{|\xi \rangle}(\alpha ,0)$ by $z$- and $x$-
  rotations.
\end{description}

Since $Q_{|\xi \rangle} ( \alpha ,\beta ) = | \langle \xi | D ( \alpha
,\beta ) | \xi \rangle|^{2}$, and using equation~(\ref{D_ksi}), one
can easily find out that the condition C1 [$Q_{| \xi \rangle}
(\alpha ,\beta ) = Q_{| \xi \rangle}( \beta ,\alpha )$]  imposes
the following restriction on $\xi$ (this is one of the possible
restrictions, but the other ones are just symmetric reflections):
\begin{equation}
  \xi = \left ( \sqrt{1+\cos^{2} \vartheta} - \cos \vartheta \right ) 
  \exp ( i \vartheta ) \, , 
  \qquad 
  -\pi/2 < \vartheta < \pi/2  \, ,
  \label{ksi}
\end{equation}
so that the $Q$-function takes the form
\begin{equation}
  Q_{| \xi \rangle} ( \alpha ,\beta ) = 
  \left ( \frac{\cos \vartheta}
    {\sqrt{1+\cos^{2} \vartheta}} 
  \right )^{2h( \alpha +\beta )} 
  \left ( \frac{\sin\vartheta}
    {\sqrt{1+\cos^{2} \vartheta}} 
  \right )^{[ h( \alpha ) +h( \beta)-h( \alpha +\beta )]} \, .  
  \label{_0012}
\end{equation}

To fulfill the condition C2, we first require that $Q_{|\xi \rangle} (
\alpha ,0) $ contains all the possible values of $Q_{| \xi \rangle} (
\alpha ,\beta )$. To this end, we note that
\begin{equation}
  \fl 
  Q_{| \xi \rangle} ( \kappa, \kappa ) = \left ( 
    \frac{\sin \vartheta}{\sqrt{1+\cos^{2} \vartheta}} 
  \right )^{2h( \kappa )} \, , 
  \qquad \qquad
  Q_{| \xi\rangle} ( \alpha , 0) = \left ( 
    \frac{\cos \vartheta}{\sqrt{1+\cos^{2}\vartheta}} 
  \right )^{2h( \alpha )} \, ,  
  \label{_0012b}
\end{equation}
so that the only possibility (apart from symmetric reflections) is
that $ \sin \vartheta =\cos \vartheta $, i.e. $\vartheta =\pi
/4$. Consequently, equation~(\ref{_0012}) takes the simple form
\begin{equation}
  Q_{| \xi \rangle} ( \alpha , \beta ) = \left ( \frac{1}{\sqrt{3}} 
  \right  )^{[ h( \alpha ) +h( \beta ) +h( \alpha +\beta )]} \, ,  
  \label{Qf}
\end{equation}
which explicitly fulfills $Q_{| \xi \rangle} ( \alpha , \alpha ) =
Q_{| \xi \rangle} ( \alpha , 0 )$.

Using the properties of the function $h$ (see appendix A), one can infer
that for any ordered pair $(\alpha ,\beta )$ there is always a field element 
$\kappa $, given by 
\begin{equation}
\kappa =\sum_{i=1}^{n}(a_{i}+b_{i}-a_{i}b_{i})\,\theta_{i}\,,
\end{equation}
$\{\theta_{i} \}$ being the self-dual basis, such that $Q_{|\xi\rangle}
(\alpha ,\beta ) = Q_{|\xi \rangle} (\kappa ,0)$. This means that each
value of $Q_{|\xi \rangle}(\alpha ,\beta )$ can be obtained by
rotating $Q_{|\xi \rangle}(\kappa ,0)$ according to the following
protocol:
\begin{equation}
  (\kappa ,0)\overset{P_{\mu}}{\mapsto}
  (\kappa ,\mu \kappa )\overset{Q_{\nu}}{\mapsto}
  (\kappa +\kappa \mu \nu ,\kappa \mu )  \, ,
  \label{PR}
\end{equation}
where the rotation parameters are $\mu =\beta \kappa^{-1}$ and $\nu
=(\alpha +\kappa )\beta^{-1}$.

To conclude, we observe that, for a single qubit, the $P$-function
(\ref{P1a}) for the value $\xi $ in equation~(\ref{ksi}) (at
$\vartheta =\pi /4$), becomes
\begin{equation}
  P(a,b) = \frac{1}{4}+\frac{\sqrt{3}}{4} 
  \left [ 
    (-1)^{b}   \Tr(\varrho \sigma _{z}) 
    + (-1)^{a} \Tr(\varrho \sigma_{x})
    + (-1)^{a+b}\Tr(\varrho \sigma_{y})
  \right ] \,.
\end{equation}
which is the most uniform possible. Note, in passing, that this
maximum uniformity could be also adopted as a reasonable criterion to
fix the value of $\xi $. The associated unit vector clearly reflects
this uniformity
\begin{equation}
  \mathbf{n} = (\langle \xi |\sigma_{x}|\xi \rangle ,
  \langle \xi |\sigma _{y}|\xi \rangle ,\langle \xi |\sigma_{z}|\xi \rangle ) 
  = 1/\sqrt{3} \,  ( 1, 1,  1 ) \, .
\end{equation}

\section{Ordering points in the discrete phase space}

The very simple form of the $Q$-function for the fiducial state $|\xi
\rangle $ in the previous section allows to introduce a partial order
in the $n$-qubit phase space. Indeed, since any $Q_{|\xi
  \rangle}(\alpha ,\beta )$ can be obtained by rotations from $Q_{|\xi
  \rangle}(\kappa ,0) = 3^{-h(\kappa)}$, we can order the points on
the horizontal axis according to the values of the $h$-function:
$0\leq h(\alpha )=k\leq n$. The $C_{k}^{n}=\frac{n!}{ k!(n-k)!}$
elements which correspond to the same value of the $h$-function remain
disordered inside the strip with the fixed value of $h(\alpha )=k$.
This automatically arranges the rest of the phase-space points
according to the symmetry property and the construction (\ref{PR}).

\begin{figure}
\includegraphics[scale=0.65]{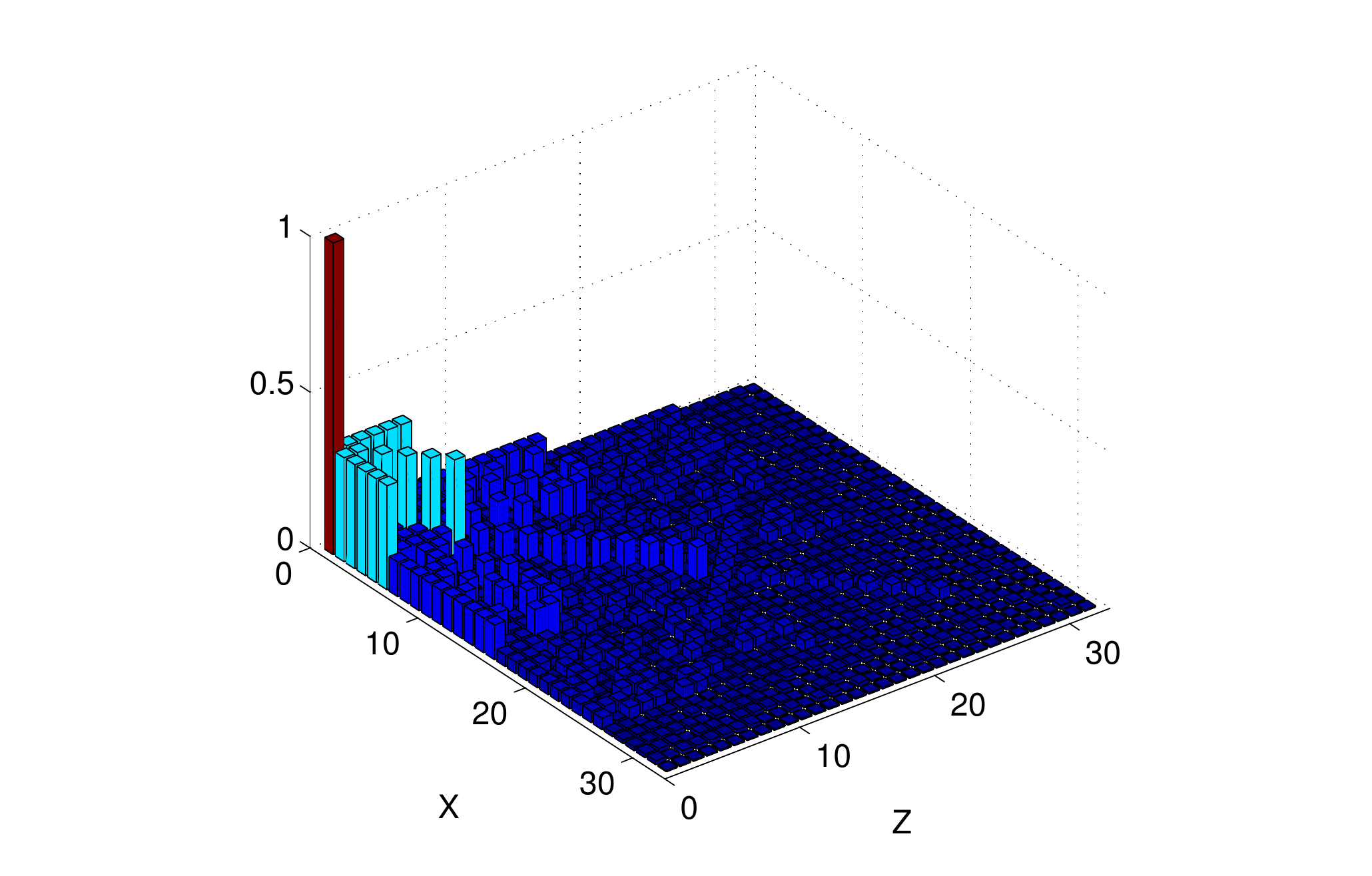} 
\caption{Ordered Q-function for the fiducial state $| \xi \rangle$.}
\end{figure}

In figure~1 we plot the $Q$-function for the fiducial state $| \xi
\rangle $ for 5 qubits using this ordering. Explicitly, the order of
axis is chosen as: $0$, $\sigma^{6}$, $\sigma^{26}$, $\sigma$,
$\sigma^{7}$, $\sigma^{2}$, $ \sigma^{16}$, $\sigma^{27}$,
$\sigma^{8}$, $\sigma^{17}$, $\sigma^{18}$, $ \sigma^{3}$,
$\sigma^{19}$, $\theta^{20}$, $\theta^{9}$, $\theta^{28}$, $
\theta^{10}$, $\sigma^{21}$, $\sigma^{22}$, $\sigma^{29}$,
$\sigma^{23}$, $ \sigma^{24}$, $\sigma^{4}$, $\sigma^{11}$,
$\sigma^{5}$, $\sigma^{12}$, $ \sigma^{25}$, $\sigma^{13}$,
$\sigma^{14}$, $\sigma^{30}$, $\sigma^{15}$, $ \sigma^{31}$. The
irreducible polynomial used is $x^{5}+x^{2}+1=0$, and the self-dual
basis chosen for $\Gal{2^{5}}$ is $\theta_{1} =\sigma^{3},
\theta_{1}=\sigma^{5},\theta_{1}=\sigma^{11},\theta_{1}
=\sigma^{22},\theta_{1}=\sigma^{24}$.

\begin{figure}[tbp]
  \begin{center}
    \includegraphics[scale=0.65]{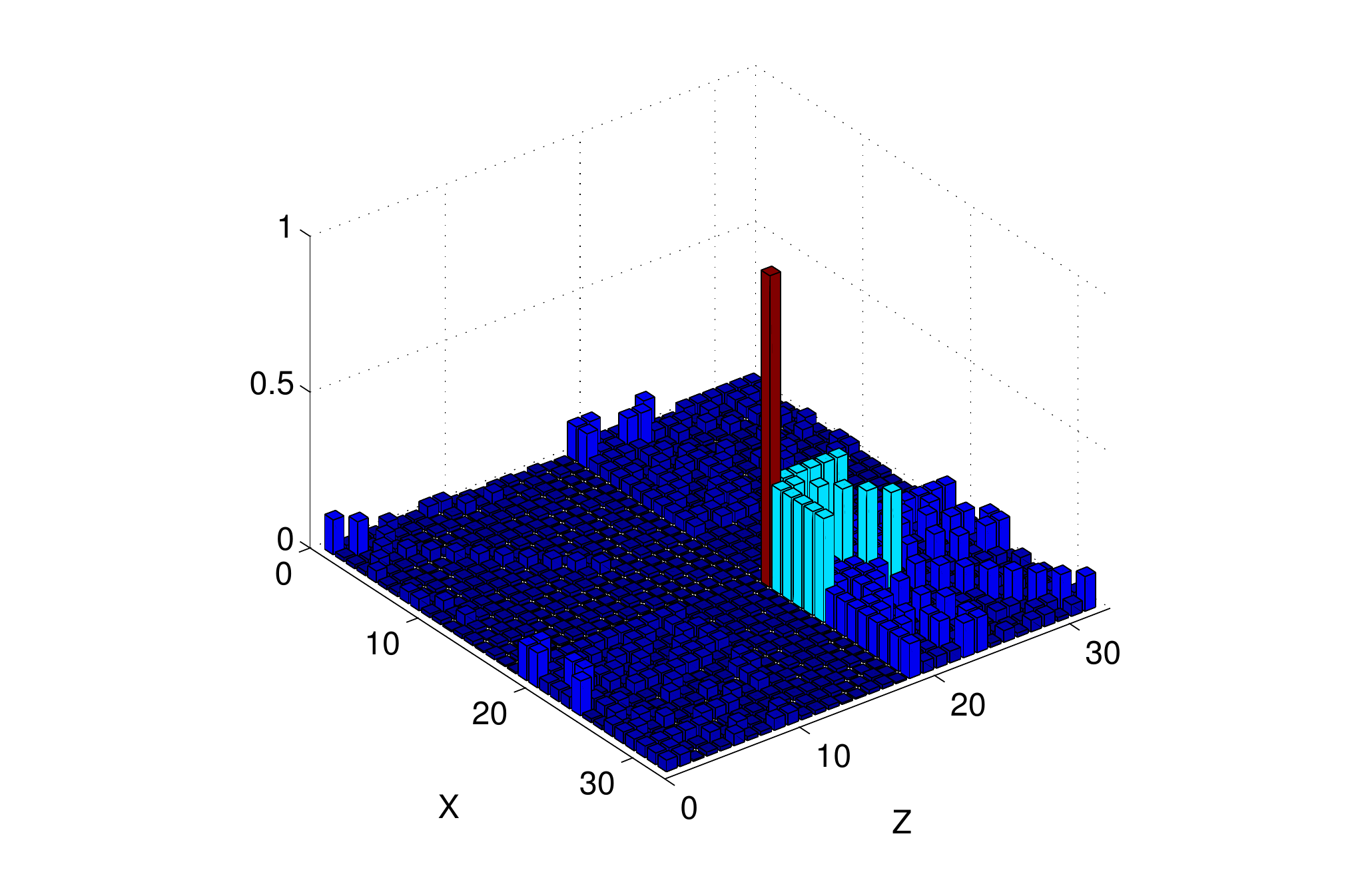}
  \end{center}
  \caption{Properly ordered Q-function for the state $|\theta^{10},
    \theta^{10} \rangle $.}
\end{figure}

The shape of the $Q$-function presents a hump localized at the
origin. Due to the covariance under displacements, $Q_{| \gamma ,
  \delta \rangle} ( \alpha , \beta ) =Q_{| \xi \rangle} ( \alpha
+\gamma ,\beta +\delta )$, the ordering should be applied to the pairs
$( \alpha +\gamma ,\beta +\delta )$, but not to $( \alpha ,\beta ) $
itself. In this sense, one cannot properly say that $Q_{|\gamma
  ,\delta \rangle} ( \alpha ,\beta ) $ has a hump located at $( \gamma
,\delta )$ if we keep the previously established order.  Nonetheless,
it is clear that due to the functional form of $Q_{| \gamma ,\delta
  \rangle} ( \alpha ,\beta ) $ the elements of the field can be easily
rearranged (using the summation table) in a such way that the
corresponding hump becomes centered at $( \gamma ,\delta ) $ and has a
symmetric form. In figure~2 we plot the $Q$-function for the - qubit
CS $| \theta^{10}, \theta^{10} \rangle = D( \theta^{10}, \theta^{10})
| \xi \rangle$ according to such a prescription.

It is also worth observing that the $Q$-function of an arbitrary state
$|\Psi \rangle $ can be written down as
\begin{equation}
  Q_{|\Psi \rangle}(\alpha ,\beta ) = \sum_{\gamma ,\delta}
  P(\alpha +\gamma,\beta +\delta )\, Q_{|\xi \rangle}(\gamma ,\delta ) \,,
\end{equation}
i.e. as an smearing of the $P$-function. In particular, the order
established by the $h$-function helps to visualize the superpositions of
several discrete CS as spatially separated humps in phase space. In
figure~3 we plot the $Q$-function for a superposition of two CS for 5
qubits, $|\Psi \rangle \propto (|\xi \rangle + D(\theta^{31},\theta^{31})|\xi \rangle )$, 
and the ordering is the same as for $Q_{|\xi \rangle}$. We can clearly
observe  two humps with a residual symmetry.

\begin{figure}[tbp]
  \begin{center}
    \includegraphics[scale=0.65]{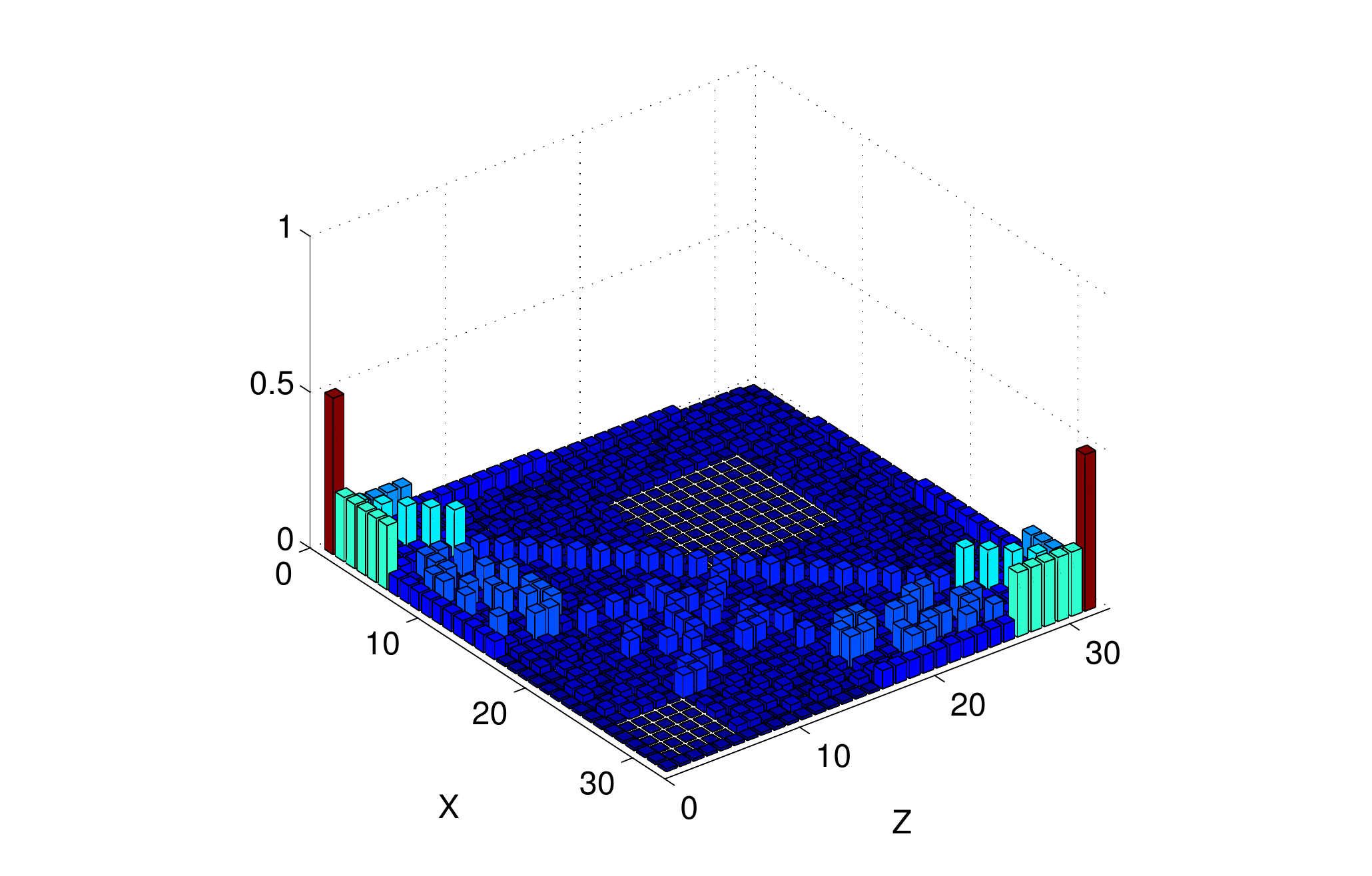}
  \end{center}
  \caption{Q-function for the state $| \psi \rangle \propto (|\xi \rangle
    + D(\theta^{31},\theta^{31}) |\xi \rangle )$.}
\end{figure}

As a final remark we may note that the distribution still can be
re-ordered in a more symmetric form just distributing the points with
the same value of $h(\alpha )$ on both sides of the principal
peak. Although the number of such points is not always even, for a
large number of qubits the distribution corresponding to the CS is
practically symmetric, as it can be seen from figure 4, where we plot
the $Q$-function for the fiducial state $|\xi \rangle$ corresponding
to 8 qubits.

\begin{figure}[tbp]
  \begin{center}
    \includegraphics[scale=0.65]{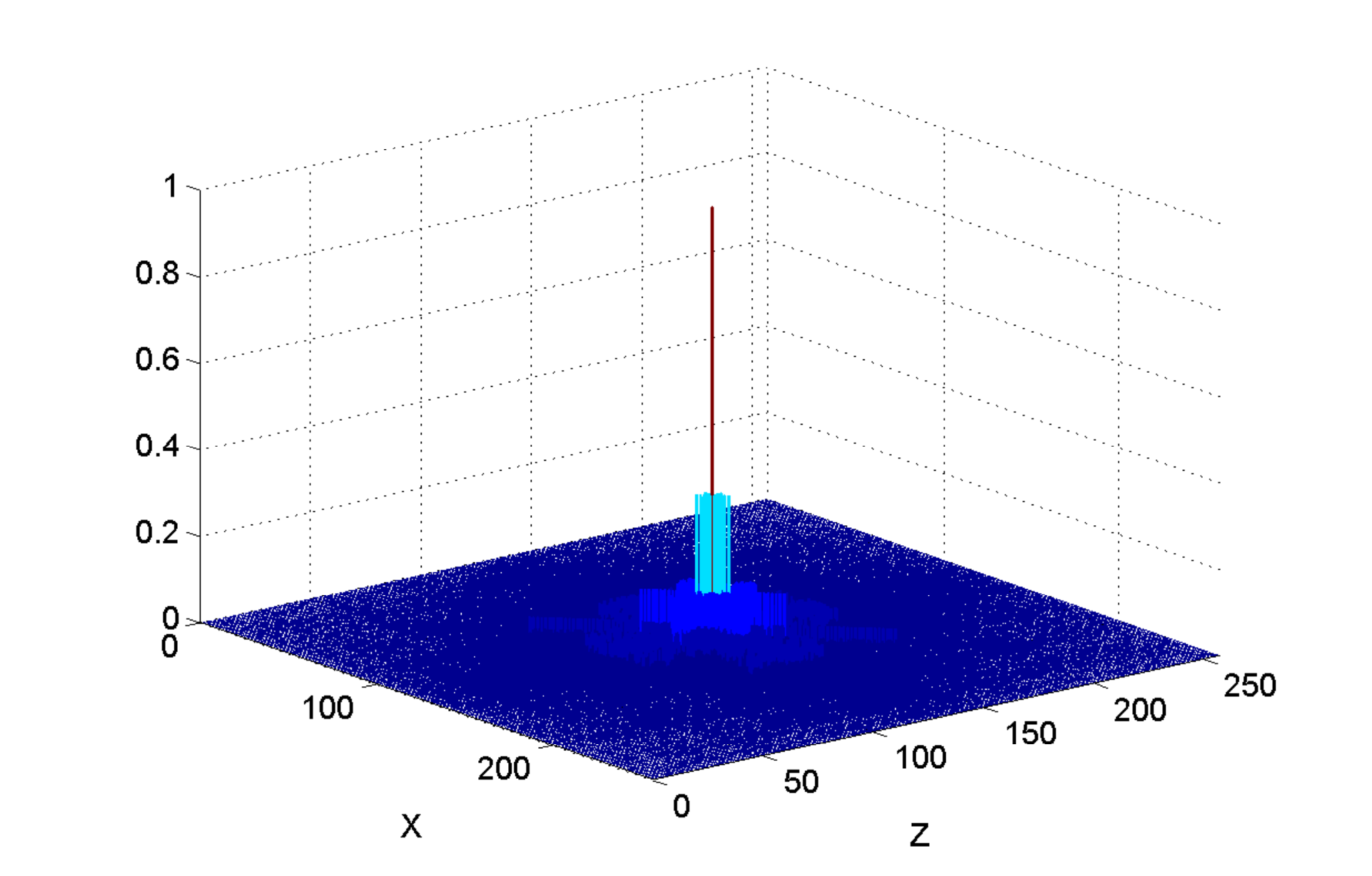}
  \end{center}
  \caption{Ordered symmetrized $Q$-function of the fiducial coherent
    state  $|\xi \rangle $ displaced to the center of the phase-space
    for eight qubits.}
\end{figure}

{It is worth comparing the form of the $Q$-function for the $n$-qubit
CS (\ref{discretecs}) in the limit $n\gg 1$ with the CS resulting from
taking as the fiducial state an eigenstate of
the discrete Fourier transform~\cite{Mehta:1987dp} . In the
latter case the $Q$-function tends to a Gaussian
shape~\cite{Galetti:1996rr} , while in our approach it has a step 
form modulated by a decreasing function $f(k)$, which along the axes 
$Z$ ($\beta =0$), $X$ ($\alpha =0$) and $Y$  ($\alpha =\beta $) has an
 exponential form:
  \begin{equation}
    f(k)=\delta_{k0} + 3^{-k}\sum_{m=0}^{n-1}
    \left [ 
    \theta \left( k-\sum_{r=0}^{m} C_{r}^{n} \right ) 
    - \theta \left( k-\sum\limits_{r=0}^{m+1} C_{r}^{n} \right)  
  \right ] \, ,
\end{equation}
where $\theta \left( k\right) $ is the Heaviside step function.

\section{Detecting correlations in $n$-qubit systems}

By construction, the discrete CS are factorized states, so the qubits
therein do not exhibit correlations. For symmetric states, the
correlations are frequently measured using the concept of spin
squeezing~\cite{Itano:1993fk,Kitagawa:1993uq,Korbicz:2006kx,
Toth:2007vn,Toth:2009ys,Ma:2012ly}, comparing the fluctuations of
some definite operator with the standard quantum limit, given by the
spin CS.  Nevertheless, for nonsymmetric states similar criteria do
not work well~\cite{Usha-Devi:2003ys} for the choice of the measured
operators becomes nontrivial.

To study correlations in nonsymmetric $n$-qubit states we apply a
criteria proposed in reference~\cite{Luis:2006zr} to quantify
polarization fluctuations. According to this approach, we compute the
sum of squares of the $Q$-function: quantum correlations make such
a sum lesser than the corresponding one for a CS. In fact, for the
fiducial state $| \xi \rangle $ we have
\begin{equation}
  \fl 
  \sum_{\alpha ,\beta} Q_{| \xi \rangle}^{2} ( \alpha ,\beta ) =
  \sum_{\alpha,\beta} 
  \left ( \frac{1}{3} \right )^{[ h( \alpha) +h( \beta )+h( \alpha +\beta )]} = 
  \prod_{i=1}^{n} \sum_{a_{i}, b_{i}= 0}^{1}
  \left ( \frac{1}{3} \right )^{[2 a_{i}+ 2 b_{i}-2 a_{i} b_{i}]} = 
  \left (  \frac{4}{3} \right )^{n} \, .  
  \label{Q2}
\end{equation}

\begin{figure}
\begin{center}
\includegraphics[scale=0.65]{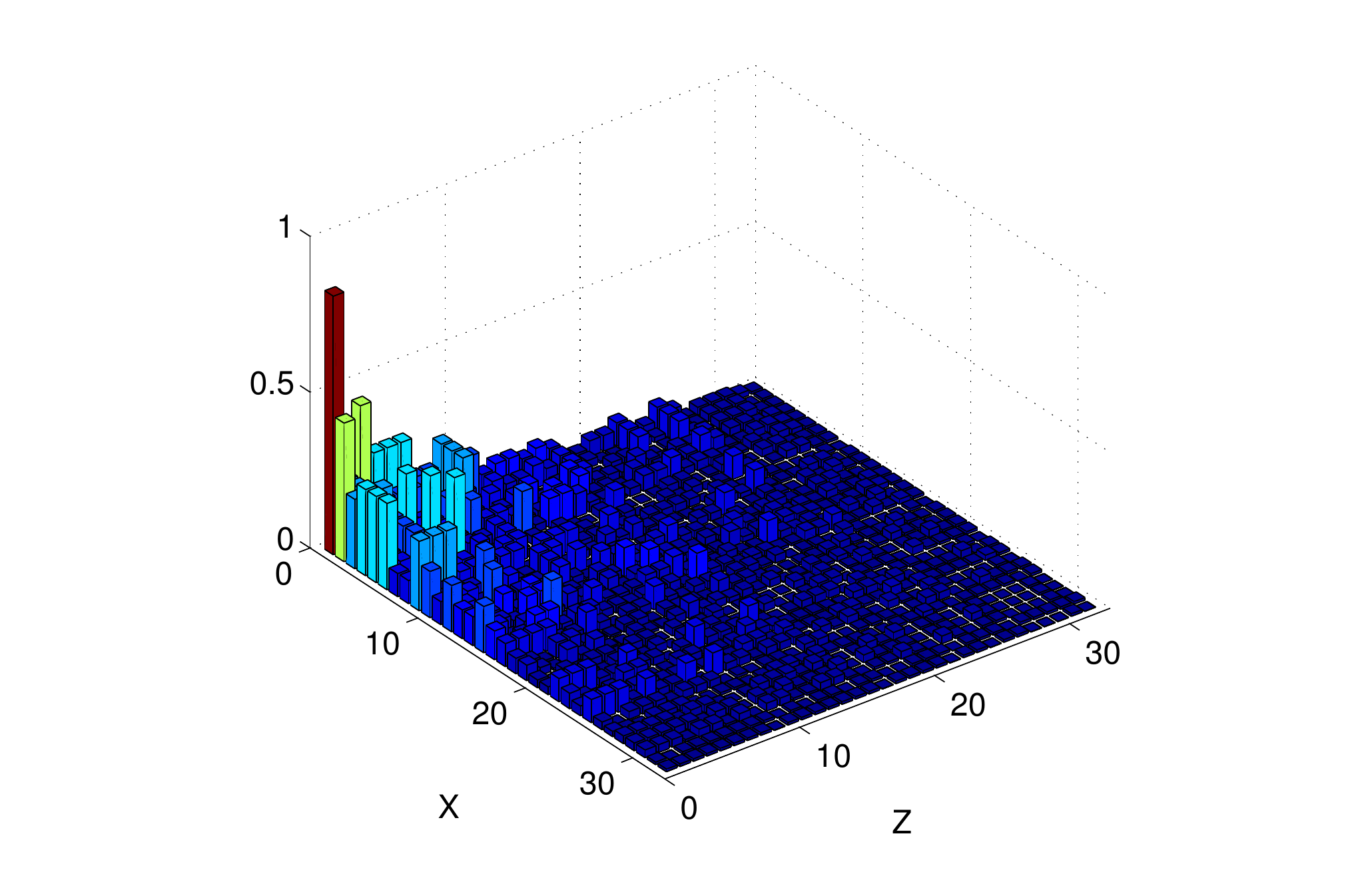}
\end{center}
\caption{Q-function for the state $\XOR_{1,2} | \xi \rangle$.}
\end{figure}

To check the method, we consider a simple way to induce correlations
between qubits: the application if $\XOR_{p,q}$ gates, where the pair
$(p,q)$ indicates the qubits on which the operator is applied, namely
\begin{equation}
  \XOR_{p,q} |a_{1}, \ldots , a_{p}, \ldots , a_{q},\ldots ,a_{n}\rangle
  = | a_{1}, \ldots , a_{p}, \ldots , a_{q} + a_{p}, \ldots ,
  a_{n}\rangle \, .
\end{equation}
For a correlated state $|\Psi \rangle =\XOR_{p,q}D(\mu ,\nu)|\xi
\rangle $ the sum of $Q^{2}$ curiously does not depend on the form of
the displacement $D(\mu ,\nu )$ and gives
\begin{equation}
  \sum_{\alpha ,\beta}Q_{|\Psi \rangle}^{2}(\alpha ,\beta )=\frac{128}{81}
  \left( \frac{4}{3}\right)^{n-2}\,,
\end{equation}
which is smaller than (\ref{Q2}). In the same vein, the application of
$k$ $\XOR$ gates between different particles (i.e., now $|\Psi \rangle
=\XOR_{p_{k},q_{k}}\ldots \XOR _{p_{1},q_{1}} D(\mu ,\nu )|\xi \rangle
$, with $p_{1}\neq q_{1}\neq \ldots \neq p_{k}\neq q_{k}$) keeps
decreasing the sum:
\begin{equation}
  \sum_{\alpha ,\beta} Q_{|\Psi \rangle}^{2} ( \alpha ,\beta ) = 
  \left (  \frac{128}{81} \right)^{k} 
  \left ( \frac{4}{3} \right)^{n-2k} \,.
\end{equation}
Similarly, the application of sequences of $\XOR$ gates to the
fiducial state also leads to decreasing values of the $\sum
Q^{2}(\alpha ,\beta )$. This effect can be clearly seen in figure~5,
where the $Q$-function for the state $\XOR_{1,2}|\xi \rangle $ is
plotted.  One can observe that the heights of the $Q(\alpha ,\beta )$
are smaller, so that the distribution initially localized at the
origin sparse over a substantial part of the phase space.

\begin{figure}[tbp]
  \begin{center}
    \includegraphics[scale=0.65]{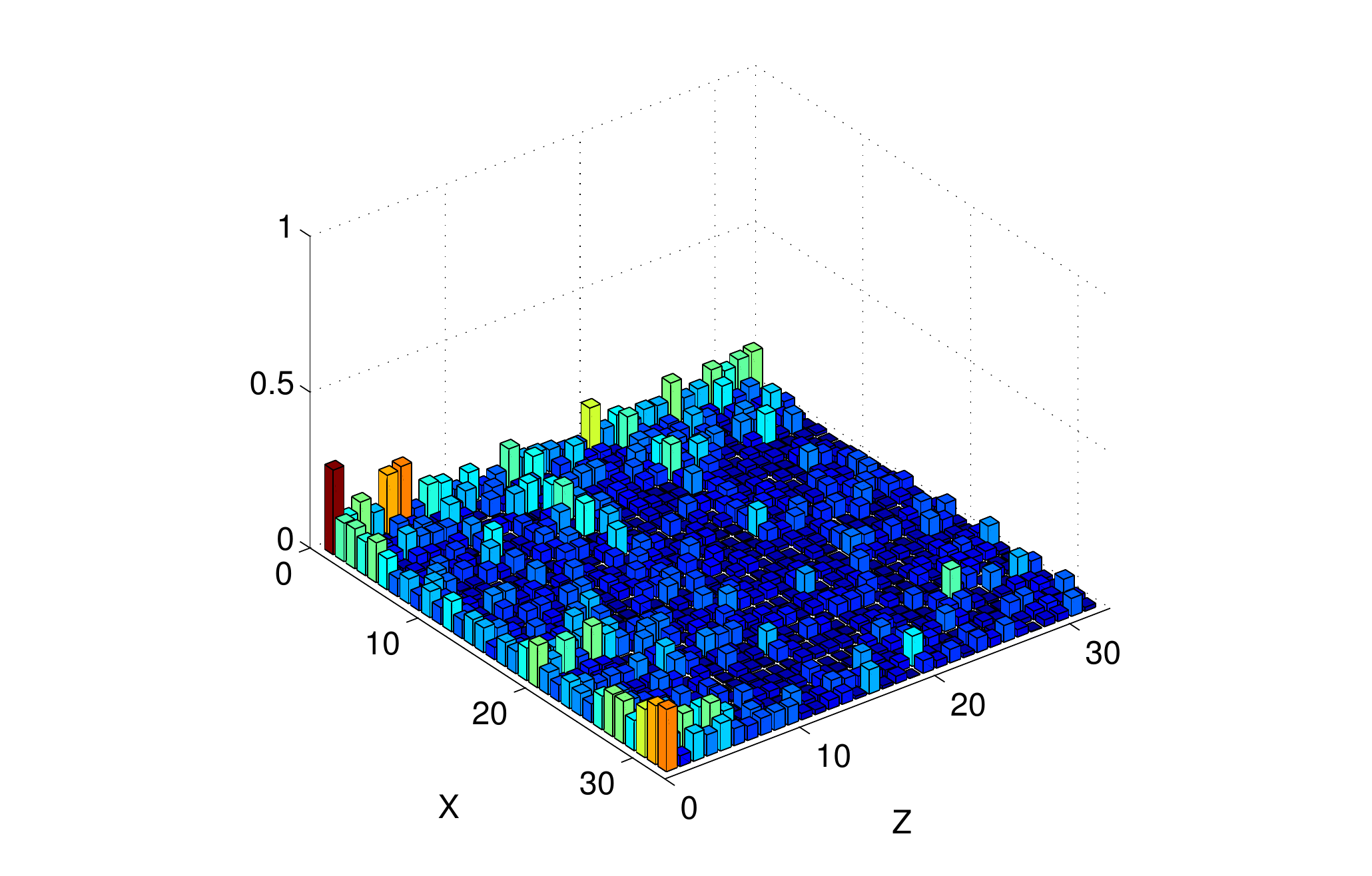}
  \end{center}
  \caption{$Q$-function of the squeezed state $S_{\sigma^{7}}|\xi
    \rangle.$}
\end{figure}

To induce correlation between all the qubits in a regular way one can
apply the squeezing operator~\cite{Vourdas:2004,Klimov:2009eu}
\begin{equation}
  S_{\zeta} = \sum_{\lambda}|\lambda \rangle 
  \langle \zeta \lambda | \,   ,
  \label{S_dn}
\end{equation}
{which acts on the $Z_{\alpha}$ and $X_{\beta}$ as the scaling transformation
  \begin{equation}
    S_{\zeta}\,X_{\beta}S_{\zeta}^{\dagger} = X_{\beta \zeta} \, ,
    \qquad \qquad
    S_{\zeta}\,Z_{\alpha}S_{\zeta}^{\dagger}\,=Z_{\alpha \zeta^{-1}} \, ,
  \end{equation}
  much as in the continuous case. The action of (\ref{S_dn}) on a CS 
  can be formally expressed as
  \begin{equation}
    S_{\zeta} | \xi \rangle =\frac{1}{(1+|\xi |^{2})^{n/2}}
    \sum_{\kappa}\xi^{h(\zeta \kappa )}|\kappa \rangle \, ,
  \end{equation}
  and implies that the initially factorized state (\ref{cs tp}) is
  transformed into
  \begin{equation}
S_{\zeta} |\xi \rangle =\frac{1}{(1+|\xi |^{2})^{n/2}}
\sum_{k_{1}, \ldots ,k_{n}} \xi^{\sum_{j=1}^{n}k_{j}d_{j}}\,|k_{1}, \ldots, k_{n}\rangle \,,
\end{equation}
where}
\begin{equation}
d_{j}=\sum_{i,m=1}^{n}f_{ijm}c_{m} \, ,
\qquad c_{m} = \tr (\zeta \theta_{m}) \, ,
\qquad
f_{ijm} = \tr(\theta_{i}\theta_{j}\theta_{m}) \,.
\end{equation}
The squeezing operator $S_{\zeta}$   correlates all the qubits in a generic
CS (\ref{discretecs}) and the degree of such correlation depends both
on $ \xi $  and $\zeta$. For example, in the 5-qubit case the
operator $\hat{S}_{\sigma^{7}}$ correlates qubits in the initial
CS with $\xi =e^{\pi i/4}\left( \sqrt{3}-1\right) /\sqrt{2}$  in
the most efficient way according to the criteria (\ref{Q2}), and its action
on the fiducial state $| \xi \rangle$ is
\begin{eqnarray}
  S_{\sigma^{7}} | \xi \rangle & = & \frac{1}{(1+|\xi|^{2})^{5/2}}
  \sum_{k_{i}} 
  \xi^{\{ k_{4}+k_{5}\}}
  \xi^{ \{ k_{2} + k_{3} + k_{4} + k_{5} \} }
  \xi^{\left\{ k_{2}+k_{4}\right\}} \nonumber \\
  & \times & \xi^{\left\{k_{1}+k_{2}+k_{3}+k_{4}\right\}}
  \xi^{\left\{ k_{1}+k_{2}+k_{5}\right\}}
  | k_{1} ,  k_{2}, k_{3}, k_{4},  k_{5}  \rangle \, , 
\end{eqnarray}
and $\{ . + . \}$ means sum $\bmod \ 2$. In figure 6 we plot the
$Q$-function of the state $\hat{S}_{\sigma^7} |\xi \rangle $, where it
can be observed that the initial distribution is spread out over
practically all the phase space.

\section{Conclusions}

We have developed a method for constructing discrete CS from the symmetry
conditions for the $Q$-function of the fiducial state. This has allowed us
to order the points in the discrete phase space. Besides, we have applied a
criterion for the detection of quantum correlations to the discrete case and
have shown that it can be useful for $n$-qubit systems.

\ack 

This work is partially supported by the Grant 106525 of CONACyT
(Mexico), the Grant PFB08024 of CONICYT (Chile), the Grants
FIS2008-04356 and FIS2011-26786 of the Spanish DGI and the UCM-BSCH
program (Grant GR-920992).

\appendix

\section{Some properties of the function $h( \alpha )$}

The function $h(\alpha )$ is defined as the number of nonzero
components in the expansion of a field element $\alpha $ in the self-dual
basis $\{\theta _{i}\}$, that is
\begin{equation}
  h(\alpha )=\sum_{i=1}^{n}a_{i} \,,
\end{equation}
where $a_{i}=\tr(\alpha \theta_{i})$. Note that $ 0 \leq h(\alpha )
\leq n$.  The basic properties we need in this paper are the
following:
\begin{eqnarray}
  \sum_{i=1}^{n}\chi (\alpha \sigma_{i})
  &=&\sum_{i=1}^{n}(-1)^{a_{i}}=n-2h(\alpha )\,,  \nonumber  \\
  & & \\
  h(\alpha +\beta ) &=&h(\alpha )+h(\beta
  )-2\sum_{i=1}^{n}a_{i}b_{i}\, , \nonumber
\end{eqnarray}
where $b_{i}=\tr( \beta \theta_{i})$. The second of these equations follows
from the equality 
\begin{equation}
h(\alpha +\beta )=\{a_{1}+b_{1}\}+\{a_{2}+b_{2}\}+\ldots +\{a_{n}+b_{n}\}\,.
\end{equation}
Here $\{\cdot + \cdot \}$ denotes again the sum $\bmod\,2$ and verifies 
\begin{equation}
\{a_{i}+b_{i}\}=a_{i}+b_{i}-2a_{i}b_{i}\,.
\end{equation}

\newpage


\providecommand{\newblock}{}

\end{document}